\documentclass[a4paper,twocolumn]{article}

\usepackage{microtype}
\usepackage{hyperref}
\usepackage{booktabs, array, multirow}
\usepackage{mathtools, amsmath}
\usepackage{amssymb}
\usepackage{bm, upgreek}
\usepackage{siunitx}
\usepackage{graphicx, caption}
\usepackage{subcaption}
\usepackage{graphbox}
\usepackage{xspace}
\usepackage{acro,pdfcomment}
\usepackage[backend = biber, style = authoryear]{biblatex}
\usepackage{upquote}
\usepackage{tikz,transparent}
\usepackage{relsize}
\usepackage{fnpct}

\hypersetup{
    pdfpagemode = UseNone,
    pdfkeywords = {},
    pdfstartview = FitH,
    colorlinks = true,
}

\graphicspath{ {./fig/} }

\setfnpct{
  after-punct-space = -0.215em
}

\newcommand{\Logitech}{Logitech\(^\text{\circledR}\)\xspace}
\newcommand{\Spotlight}{Spotlight\(^\text{\texttrademark}\)\xspace}
\newcommand{\LogiOptionsPlus}{Logi~Options\hspace{-0.05em}\raisebox{0.4ex}{\relsize{-3}+}\xspace}

\renewcommand{\bold}[1]{{\textbf{#1}}}

\title{Case study of a national-level academic conference organised in hybrid mode at low cost}
\author{
    Violet M. Harvey\(^{1}\)\thanks{Email: violet.harvey@adelaide.edu.au}
    \and Simon Lee\(^{1}\)
    \and Bruce Dawson\(^{1}\)
    \and Sabrina Einecke\(^{1}\)
    \and Gavin Rowell\(^{1}\)
    \\
    \makebox[0pt]{\parbox{\textwidth}{%
        \vspace{0.5em}\relsize{-1}%
        \({}^{1}\)School of Physics, Chemistry and Earth Sciences, University of Adelaide, Adelaide SA 5005, Australia
    }}
}
\date{6 March 2026}

\makeatletter
\hypersetup{
    pdftitle  = {\@title},
    pdfauthor = {Violet M. Harvey et al.},
}
\makeatother

\begin{document}

\maketitle

\section{Introduction}

    In July 2025, the University of Adelaide hosted the Astronomical Society of Australia's Annual Scientific Meeting on its North Terrace campus\footnote{
        The conference program is hosted at \url{https://indico.global/e/asm2025}.
    }.
    We ran the conference in a hybrid mode, with options for in-person and online attendance.
    The conference ran for five days, Monday to Friday, with sessions starting at 10:00 ACST and finishing between 15:00 and 18:00 ACST (depending on the day).
    There were two parallel streams in the morning, a plenary stream through the middle of the day, and two parallel streams again in the afternoon. 
    This report details the procedures that we used to enable the online mode of the conference at minimal cost and minimal inconvenience to the in-person attendees.

    This report is organised as follows.
    Section~\ref{sec:objectives} lays out our objectives for the hybrid organisational approach.
    In Sections~\ref{sec:hardware} and~\ref{sec:software} we discuss our choices of hardware and software, respectively, and how we integrated these systems together.
    Section~\ref{sec:procedures} summarises our experience of organising a local AV~team and the procedures that we set for running the AV in each session.
    In Section~\ref{sec:stats} we present statistics of the online attendance numbers and post-conference survey feedback.
    In Section~\ref{sec:lessons} we discuss the lessons we feel other organisers may particularly be able to learn from and comment on some of the understated benefits of enabling an online component for a chiefly in-person conference.
    Section~\ref{sec:conclusion} presents our conclusions.

\begin{figure*}
    \centering
    \includegraphics[width = 0.8\textwidth]{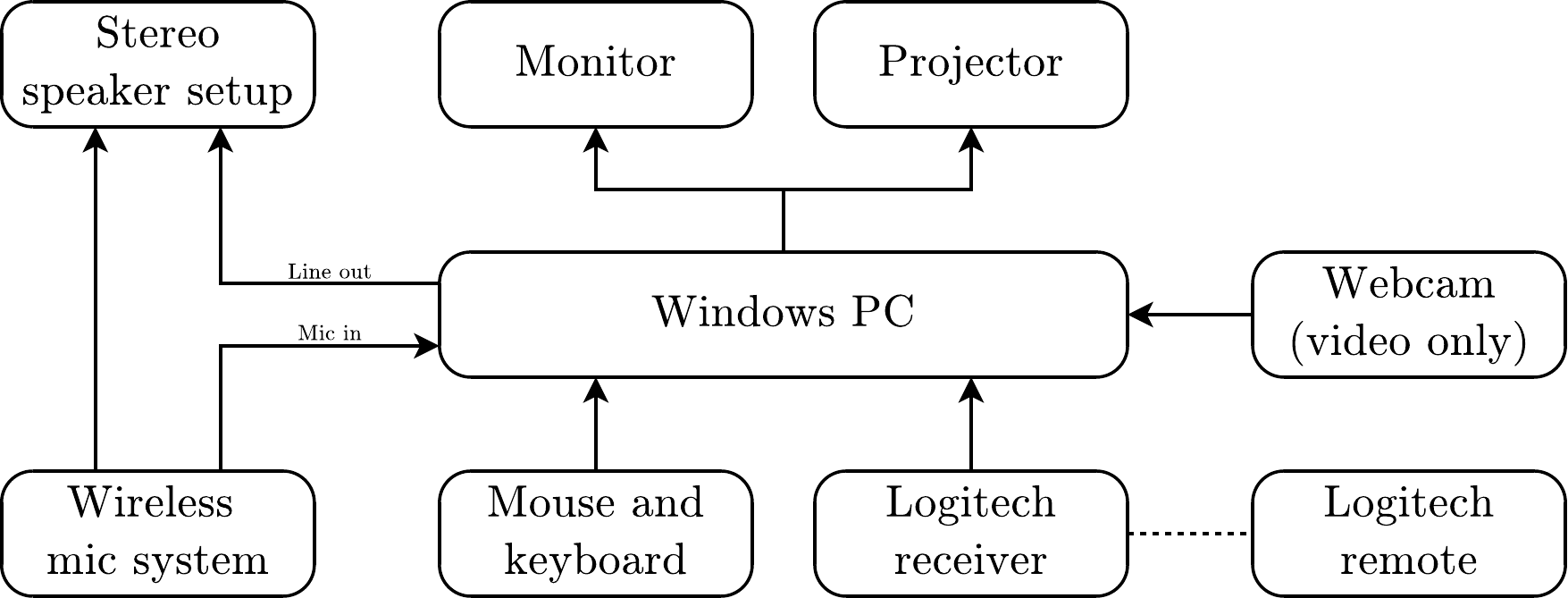}
    \caption{%
        Flowchart illustrating the hardware AV components and connections.
        It should be noted that this setup was designed around the equipment already available in each lecture theatre.
    }
    \label{fig:hardware:flowchart}
\end{figure*}

\section{Objectives}
\label{sec:objectives}

    We organised the conference with an in-person-first, online-second priority, but we desired for online participants to still have the ability to interact with speakers and ask questions.
    Secondarily, we wanted speakers to be able to present their talk remotely with an online registration.
    We prioritised ensuring that oral sessions were made available for online attendees, but we did not plan any interactive experience for the poster viewing sessions or any online networking opportunities, due to a lack of resources to manage this.
    Under the hybrid conference framework of \textcite{Moss2025} this approach lies somewhere between Level~2 and Level~3, as the online component is not fully passive, but neither is it fully integrated with its own activities.
    Despite the lack of a dedicated online poster experience, online attendees could still view the posters online asynchronously, as we requested that all poster presenters upload a PDF of their poster to the conference platform, and we included the poster ``sparkler'' sessions\footnote{
        The sparklers consisted of a 1-minute time slot during the plenary sessions for each poster presenter to use as they saw fit.
        We permitted each sparkler to be accompanied by a single slide shown whilst the speaker was on stage or live in Zoom, or a pre-recorded video.
    } in the general oral sessions available to online attendees.
    Finally, we wanted to keep costs low in the pursuit of these objectives, and prioritised the integration of our procedures with existing university infrastructure wherever possible.

\section{Hardware}
\label{sec:hardware}

    Figure~\ref{fig:hardware:flowchart} illustrates the hardware AV components and how they connected together.
    At a broad level this is the same setup for both of our conference venues, though some hardware details differed.
    This section begins with an overview of the venues we used to host the conference talks and the AV equipment already available in those rooms.
    This is followed by short discussions of the remaining equipment which was either hired, loaned, or purchased.
    See Table~\ref{tab:table:costs} for a summary of all AV-related expenses.

    \begin{table*}
        \centering
        \caption{
            Summary of AV-related expenses.
            All listed costs are inclusive of GST where applicable.
        }
        \label{tab:table:costs}
        \begin{tabular}{l S l}
            \toprule
            Item & Cost~{(AUD)} & Primary reason \\
            \midrule
            \parbox{0.3\textwidth}{Wireless microphones, receivers, audio mixer, labour} & 1887.75 & \parbox{0.45\textwidth}{In-person (to have sufficient audience microphones in the second venue)} \\
            Presentation remotes (\(2\times\)) & 358.00 & \parbox{0.45\textwidth}{Online (to see the laser pointer over Zoom)} \\
            Honorariums for AV volunteers & 1400.00 & In-person \\
            Honorariums for mic runners & 275.00 & In-person \\
            \midrule
            \textbf{Total} & 3920.75 & \\
            \bottomrule
        \end{tabular}
    \end{table*}

    \subsection{Venues and university-provided AV equipment}

        By hosting the conference on the university campus we had access to university lecture theatres.
        We used two in particular; a large theatre that hosted the plenary sessions and some parallel sessions, and a smaller nearby theatre that only hosted parallel sessions.
        Each lecture theatre was equipped with a university-provided \bold{Windows PC} connected to a \bold{monitor} and \bold{projector} (mirrored displays), and a \bold{wireless microphone system} connected to a \bold{stereo speaker setup}.
        The audio output of the Windows PC was also connected to the stereo speaker setup.
        The wireless microphone system consisted of one lapel mic appropriate for a presenter, and one or more hand-held mics appropriate for being run to audience members during question time.

        We had two essential requirements of the microphone system that we would use during the conference.
        The first requirement was that the microphone audio was routed to both the room speakers as well as to the PC as a microphone input device.
        This was to ensure we could provide a clear audio feed for the online attendees.
        The second requirement was that there were a minimum of four microphones -- one for the speaker, one for the session convener, and two for audience questions.

    \subsection{Equipment that was hired or loaned}

        We acquired some high-cost equipment for temporary use through formal hire arrangements for a price or informal loans for free.

        \subsubsection{Microphone system}

            In one lecture theatre, the university-provided microphone system met our requirements and we were able to use it seamlessly for our purposes.
            In the other lecture theatre, we found that the provided microphone system failed to meet both requirements.
            The provided microphone system was not routed to the university-provided PC (instead, it connected directly to the stereo speaker setup only), and there were only two wireless microphones provided.
            The university's AV support team advised us that it was not possible for them to connect additional wireless microphones to the room, even on a temporary basis, because of fundamental constraints with the existing audio mixer.
            They also implied it would be difficult to connect the microphone audio to the PC.
            Through helpful discussions with the university's AV support team we established the particular pieces of equipment that we needed to request from a third-party AV company such that they could be integrated with the existing equipment in the room.
            We hired four wireless microphones, wireless receivers, an audio mixer, and labour to set up the equipment and pack it away again after the conference.

            The hiring of this equipment, at \num{1887.75}~AUD (incl. GST), represents the single largest cost related to AV for the entire conference.
            However, it is important to stress that this cost was not specifically to enable the online component of the conference -- it was also necessary in order to have the minimum of four microphones in this venue, as outlined in our requirements.
            This cost would have been borne even if the conference had no online provisions.

        \subsubsection{Webcams}

            As shown in Figure~\ref{fig:hardware:flowchart} we had a \bold{webcam} attached to each PC.
            The video feeds from the webcams were for the benefit of online attendees.
            We loaned two webcams from students in the astrophysics department, two USB extension cables (one from a student, one from physics lab staff), and two tripods (both from a member of staff passionate about astrophotography).
            As such, there were no costs associated with any of this equipment.
            This was enough equipment to situate the webcams a reasonable distance away from the PC and effectively frame up the speaker. 

    \subsection{Equipment that was purchased}

        We acquired some low-cost equipment by purchasing it with conference operational funds.
        Specifically, we purchased two \bold{\Logitech \Spotlight Presentation Remotes} at a total cost of \num{358}~AUD.
        The remotes could connect to the PC via Bluetooth or a dedicated wireless receiver, and for reliability we used the dedicated receiver.
        We used these presentation remotes to operate a ``virtual laser pointer'', which is detailed in the software discussion below.
        We made the choice to use a presentation remote with a virtual laser pointer, rather than a physical laser pointer, specifically for the benefit of online attendees to be able to see what the speaker was indicating on each slide.

\section{Software and web platforms}
\label{sec:software}

    Here we list the software and web platforms that we used for the conference, and discuss relevant aspects of their configuration and integration with the hardware.

    \subsection{Zoom}

        At the core of our hybrid approach was the \href{https://zoom.us}{Zoom} videoconferencing software.
        We ran Zoom on the venue PC, which was connected to the projector as well as the audio feed from the microphone system.
        We used the Zoom screen share to capture the speaker's slides (which we required to be uploaded to the conference platform before each session started).
        We chose Zoom because of our existing institutional educational license for the software.
        This license granted us the ability to host meetings of unlimited duration with up to 300 attendees.
        Furthermore, for the smoothest integration with the venue AV equipment we aimed to run all presentations and audio through the university-managed PC, and Zoom was already installed on these PCs per university IT policy.

        We tried to maintain a straightforward experience for online attendees.
        To this end, the majority of the sessions were hosted in two Zoom rooms, which were simply named for the physical venue that they corresponded to.
        This meant that the appropriate Zoom room to join for a given session was easy to determine from the conference timetable, and in most cases there were only two in question.
        However, we had a policy that the Zoom connections for the talk sessions were restricted to conference attendees only and these connection details had to be protected.
        A handful of sessions were shared with the wider ASA, specifically the Chapter meetings, the AGM and Decadal Plan session, and the Harley~Wood Lecture.
        Therefore, we also had three additional Zoom rooms for handling these ``public'' events, so that the connection details for these Zoom rooms could be shared more widely without compromising the regular session connection details.

    \subsection{\texorpdfstring{\LogiOptionsPlus}{Logi Options+}}

        In order to enable the ``virtual laser pointer'' feature of the \Logitech presentation remotes, it was necessary to install \href{https://www.logitech.com/en-au/software/logi-options-plus.html}{\LogiOptionsPlus~(Plus)} on the university-managed Windows PCs.
        This required coordination with university IT staff and was not without issue (see further discussion in Section~\ref{sec:lessons:institutional}).
        Without this software installed, the presentation remotes could move the mouse cursor around but there was no ``dot'' that highlighted the position of the cursor.
        With this software, once correctly configured, a large dot is drawn around the cursor for as long as the spotlight button on the remote is held down.
        \LogiOptionsPlus allows this dot to be customised in size and colour, and configured to be hidden when idle so that no part of the slides is obstructed when the pointer is not active.

    \subsection{YouTube}

        We made use of the capability for Zoom to be livestreamed to \href{https://www.youtube.com}{YouTube}, with a feed on the official ASA channel.
        Livestream integration is a built-in feature of Zoom that automatically mixes the active screen share and video feed together.
        Instructions from Zoom for configuring this feature are given \href{https://support.zoom.com/hc/en/article?id=zm_kb&sysparm_article=KB0062284}{here}.
        To have control over the properties of the livestream and run multiple livestreams in parallel, we used what Zoom calls the ``stream to a YouTube Event'' configuration.
        This is achieved by enabling ``Custom Live Streaming'' for each Zoom meeting room, and linking each room to a unique ``stream key'' set up in YouTube Studio.
        As we had at most two YouTube streams running in parallel, we only needed two stream keys.
        To reduce moderation overhead, we disabled all comment and chat functionality on YouTube.
        The YouTube stream links were distributed to the full ASA membership, whilst the Zoom meetings were generally only accessible to conference registrants (with exceptions for a few sessions as previously noted).

        YouTube livestreams automatically turn into published YouTube videos after the stream concludes, which conveniently gave us our recordings of each day.
        We found that whenever Zoom was disconnected from the YouTube stream (as during breaks), as long as the stream had not been ``stopped'' in YouTube Studio, then when Zoom was reconnected to the stream, the period of missing video was snipped from the recording.
        After the livestream was concluded each day, we manually added chapter markers to each published video to make it easier for viewers to find specific sessions.

    \subsection{Slack}

        We created dedicated discussion channels on the official ASA \href{https://slack.com}{Slack} instance.
        We used one channel per physical venue, for discussion of whatever sessions occurred in that venue.
        The ASA~Slack contains a large number of the ASA members, however we expected that in-person attendees would not make much use of it during the conference.
        We did not promote Slack very heavily, and likely as a consequence of this almost no discussion of the conference talks ever took place in Slack.
        However, it did still serve as a useful centralised ``discussion forum'', and on the first day of the conference was used by some in-person attendees to discuss access to the university wi-fi.

    \subsection{Indico}
    \label{sec:software:indico}

        We used \href{https://indico.global}{Indico} as the conference platform.
        In particular we used the Indico~Global instance, which is freely available for organising not-for-profit scientific events with an attendance of less than 250 people.
        Many

        of Australia's largest institutions are connected to the international \bold{eduGAIN} federation service via subscription to the Australian Access Federation, and Indico Global permits logging in with eduGAIN.
        This meant that the majority of attendees with institutional email accounts were able to sign~in to Indico without needing to set their own username and password, significantly reducing friction to accessing secure content on the conference platform.

        We required each speaker to upload their slides to Indico before their session started.
        An Indico feature that proved very useful for the AV team to bring up each speaker's slides in the sessions was the ``material package'' download.
        For a selected session, this feature prepared a ZIP file containing the materials for each contribution in the session in individual folders named for the scheduled start time of that contribution.
        This meant all the materials for a session could be downloaded once at the start of the session, and it was quick and easy to navigate from one talk to the next during the session because they were sorted by start time.

        Indico is familiar software to members of the particle physics and astroparticle physics communities in Australia, however we found it was new to some members of the astronomical community.
        Some attendees reported issues with understanding the interface and navigating to find specific contribution details.
        We expect these to be teething issues, and would still encourage future organisers of this conference to consider using Indico.
        The tools it provides to organisers, such as inbuilt reviewing of abstracts on specified ``tracks'', are valuable, and the software is and always will be free to use thanks to its open source development by CERN.

        \subsubsection{Custom tools}

            VMH wrote two small Python tools for automating aspects of the conference platform, which can be found \href{https://github.com/vmharvey/indico-tools}{here}.
            The first tool used the Indico API to ``protect'' uploaded materials (slides and posters) and restrict their access to logged-in registrants only.
            We found this was necessary because we wanted the timetable and contribution details to be publicly viewable, which made the inherited visibility of contribution materials public as well.
            The Indico management area did not offer a more fine-grained way to set the default access level for uploaded materials to private while keeping the contribution details public.
            By periodically running this script, any materials uploaded since the last time the script was run would have their permissions updated.
            VMH configured a cron~job on a local server to run the script every two minutes.
            
            The second tool used the Indico API to read the conference timetable and a Slack webhook to post an announcement of the title and speaker for each talk in each session in the appropriate Slack channel.
            The announcements were primarily intended to contextualise the questions posed by online attendees by demarcating the period of each talk, though as noted earlier the takeup of Slack for this use was low.
            Still, for those who had Slack open it served as a simple reminder of the timetable without needing to navigate it through Indico.
            This tool was also kept running on a local server.

\section{Student volunteers and AV procedures}
\label{sec:procedures}

    We sought volunteers from two student cohorts around the physics department to help with running technical aspects of the conference.
    We assigned HDR volunteers to the AV~team, and undergraduate volunteers to the separate microphone runner team.
    These teams and their roles are described here, in context with the procedures they followed during sessions.

    \subsection{AV~team}
        
        The AV~team was comprised of two coordinators~(academic staff; VMH and SL) and 15~HDR~volunteers.
        These HDR~volunteers only contributed to conference organisation through their AV~duties, though many of them also had posters or talks at the conference in a student capacity.
        Collectively, the AV~team performed 92.5 person-hours of work over the week of the conference, contributing an average of 0.15 FTE per person.

        At the conclusion of the conference, in consideration of available funds, volunteers were awarded an honorarium (in the form of a gift card) commensurate with the amount of time they volunteered, at a rate of 15~AUD per hour.

        Volunteers were also invited to partake of the catered breaks and lunches on the days that they worked (though this was only relevant to individuals who otherwise were not attending the conference in a student capacity).

        Ahead of the conference, the coordinators spent several weeks refining a step-by-step checklist of how to run the AV for each session, from logging in and connecting to Zoom, downloading the session slides, starting the YouTube livestream, and setting up Adobe Acrobat and Microsoft PowerPoint.
        They also prepared instructions for configuring the presentation remotes, the room lighting, and other odd jobs that fell to the AV~team.
        The coordinators ran two one-hour training workshops the week before the conference to instruct the volunteers on all of these procedures and provide an opportunity to practice the steps with Zoom and YouTube.
        The time that volunteers spent at the training workshops was also considered in the value of their honorarium.

    \subsection{Microphone runners}
        
        Separate to the AV~team, there was a ``microphone runner'' team comprised of one coordinator~(academic staff; BD) and 11~undergraduate~volunteers.
        Each session was assigned two microphone runners, who had the job of delivering a hand-held microphone to audience members who raised their hands to ask questions after each talk.
        Collectively, the microphone runner team performed 70 person-hours of work over the week of the conference, making for an average of 0.17 FTE per person.
        Volunteers were invited to partake of the catered breaks and lunches on the days that they worked and awarded an honorarium (in the form of a gift card) with a flat amount of 25~AUD each.
        We considered the microphones and microphone runners essential for the in-person experience as much as the online experience, as the microphone feed was routed to both the venue speaker system and the Zoom meeting.

    \subsection{Assignment of volunteers to sessions and summary of procedures}
    
        Most sessions were assigned two AV~helpers and two microphone runners.
        The two AV~coordinators took shifts on AV~help just the same as the HDR~volunteers.
        The only exception to usual shift assignments were the Chapter sessions, which were assigned presentation support but no question support or microphone runners.
        This arrangement was made because the Chapter sessions were informally organised and not required to conform to our hybrid policies (i.e., live streaming of the session and enabling questions from online attendees).

        There were two roles for the AV~helpers, which we named ``presentation support'' and ``question support''.
        Sessions with two AV~helpers would have one to fill each of these roles.
        The presentation support person had the jobs of swapping between speakers' presentations on the projector and managing the Zoom audio and screen share.
        The question support person had the job of watching for questions posed through the Zoom chat and Slack channels (to be brought to the attention of the session convener, a separate role managed by the SOC), and some miscellaneous jobs such as helping each speaker with the lapel microphone and presentation remote.
        The question support person used their personal laptop for monitoring of Zoom and Slack, as the venue PC was dedicated to showing the slides and running the screen share.
        Whilst the presentation support person was essential to both the in-person experience of the conference as well as the online experience (as they had charge of both the local projection and the screen share), the question support person was mostly concerned with the online experience.
        However, we did find it useful that the question support person was able to take some in-person tasks off of the presentation support person by helping each speaker with the microphone and presentation remote.

        In general, the AV~procedures were in service of ensuring that all presentation materials and audio went through the venue PC and thus through the Zoom meeting hosted on that PC.
        Speakers were required to upload their presentation materials to Indico at least 30~minutes before the session began.
        We required that a PDF version of the slides was always uploaded, and optionally a PowerPoint~(PPTX) could be provided as well.
        If a PowerPoint was uploaded then the AV policy was to use the PowerPoint version of the slides, except in cases of technical issues in which case the PDF was used as a fallback.
        Presentation support began each session by downloading the Indico material package for the session (see Section~\ref{sec:software:indico}) and confirming that materials for each contribution were present.
        If speakers wanted to test whether their PowerPoint worked correctly, and hence whether to use it or the PDF backup, they had the opportunity to speak with the AV~team and confirm this before the session started.
        After starting the Zoom meeting and initialising the YouTube livestream connection, the presentation support person was concerned with beginning a screen share and opening each speaker's slides in fullscreen view.
        Online attendees were permitted to share their own screen to deliver their talk remotely.
        Otherwise, all presentations were delivered by either PDF or PPTX run on the venue PC.
        This policy led to a handful of issues with broken PowerPoint animations or embedded videos, and some speakers had Google~Slides which we never planned out how to accommodate for.
        With hindsight, we may have set our procedures a bit differently.
        We comment on this further in Section~\ref{sec:lessons}.
        Questions were taken from the in-person audience as well as from Zoom and Slack, with online questions being flagged to the session convener during question time, so they could be repeated to the speaker and audience.
        
\begin{figure*}[t]
    \hspace{-0.08\textwidth}
    \begin{subfigure}{0.55\textwidth}
        \includegraphics[clip, trim = {0 0 40 40}, width = \textwidth]{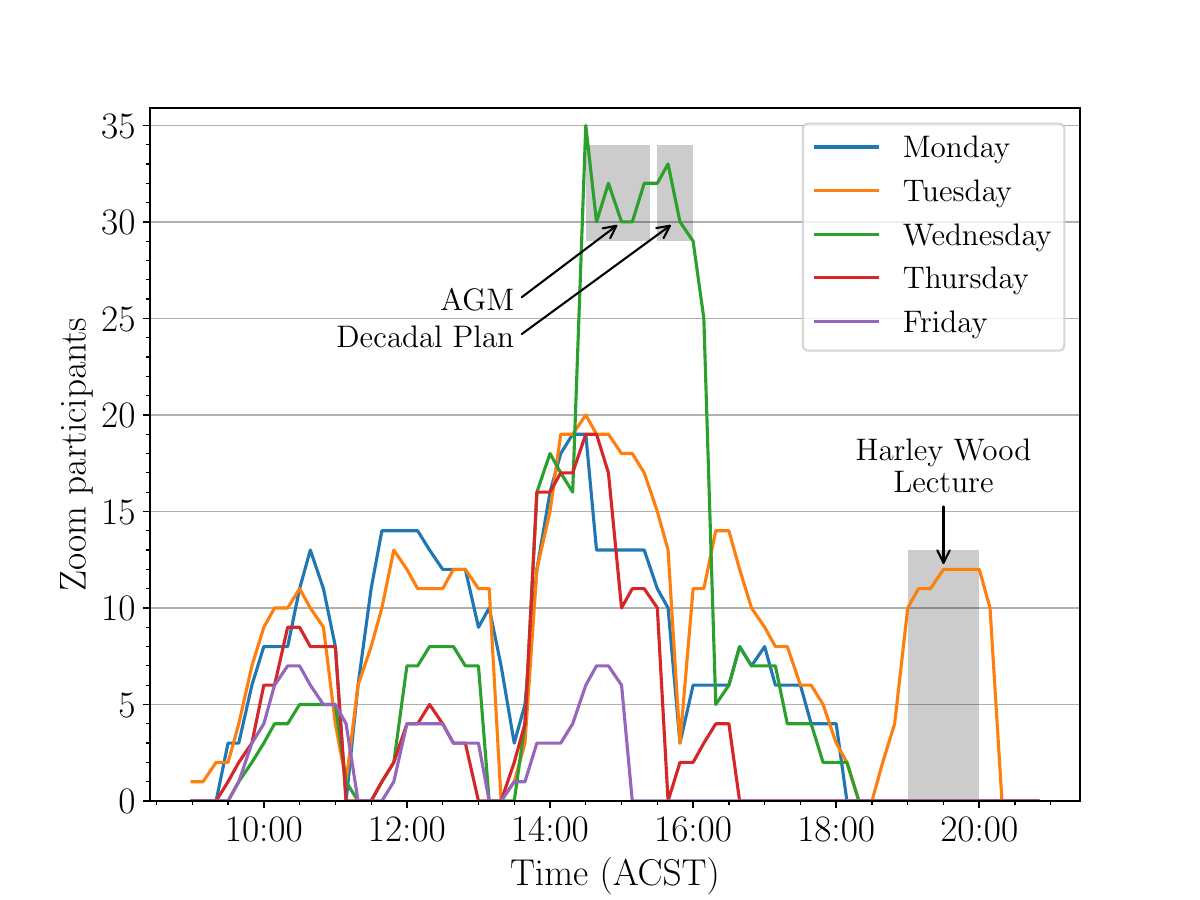}
        \caption{Zoom attendance.}
        \label{fig:stats:live:zoom}
    \end{subfigure}%
    \begin{subfigure}{0.55\textwidth}
        \includegraphics[clip, trim = {0 0 40 40}, width = \textwidth]{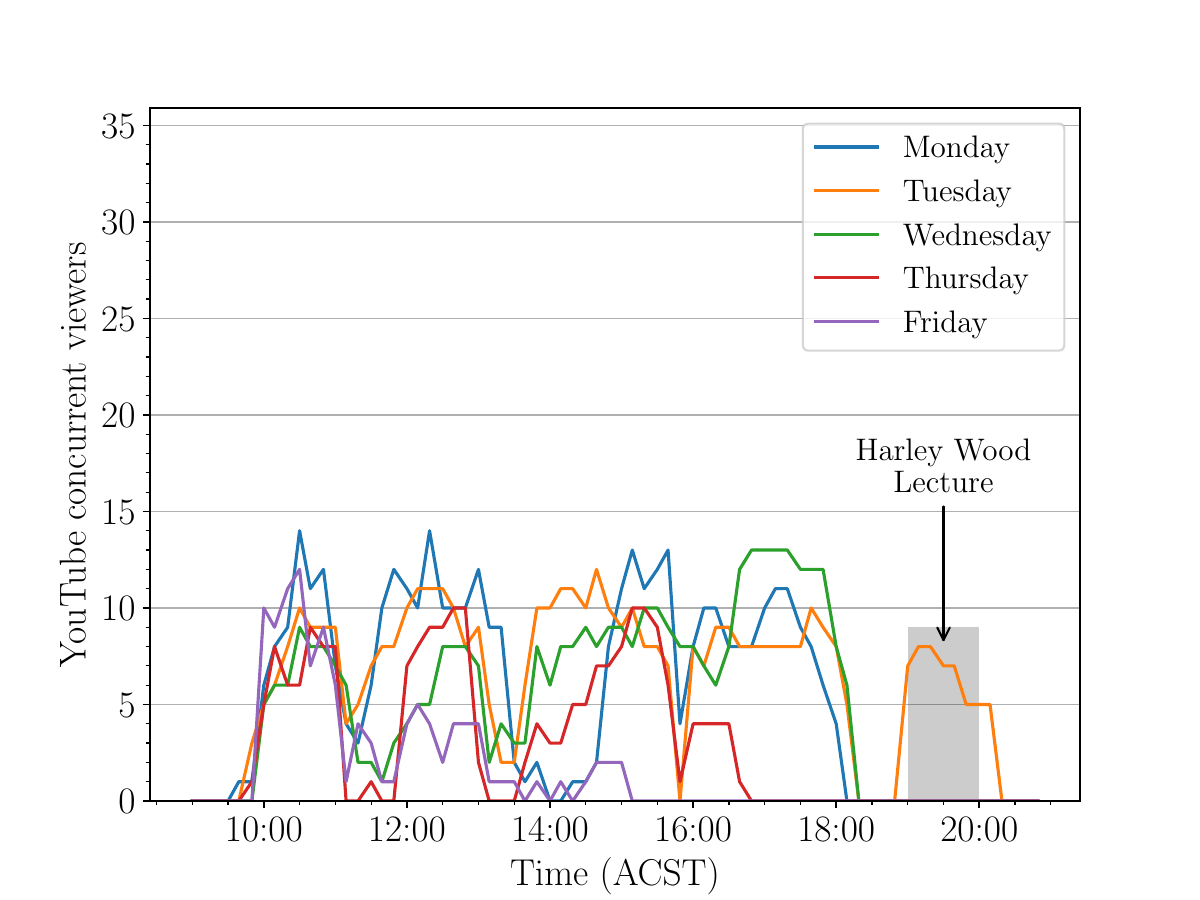}
        \caption{YouTube live viewership.}
        \label{fig:stats:live:yt}
    \end{subfigure}
    \caption{
        \emph{Left:}~Count of Zoom participants online across all active Zoom meetings each day.
        The count of Zoom participants excludes all Adelaide staff and AV~volunteers.
        \emph{Right:}~Count of concurrent YouTube viewers across all active streams each day.
        Recall that there were a total of 34 online registrants, and that the AGM, Decadal Plan, and Harley Wood Lecture were open to unregistered viewers.
    }
    \label{fig:stats:live}
\end{figure*}

\begin{figure*}[t]
    \hspace{-0.08\textwidth}
    \begin{subfigure}{0.55\textwidth}
        \includegraphics[clip, trim = {0 0 40 40}, width = \textwidth]{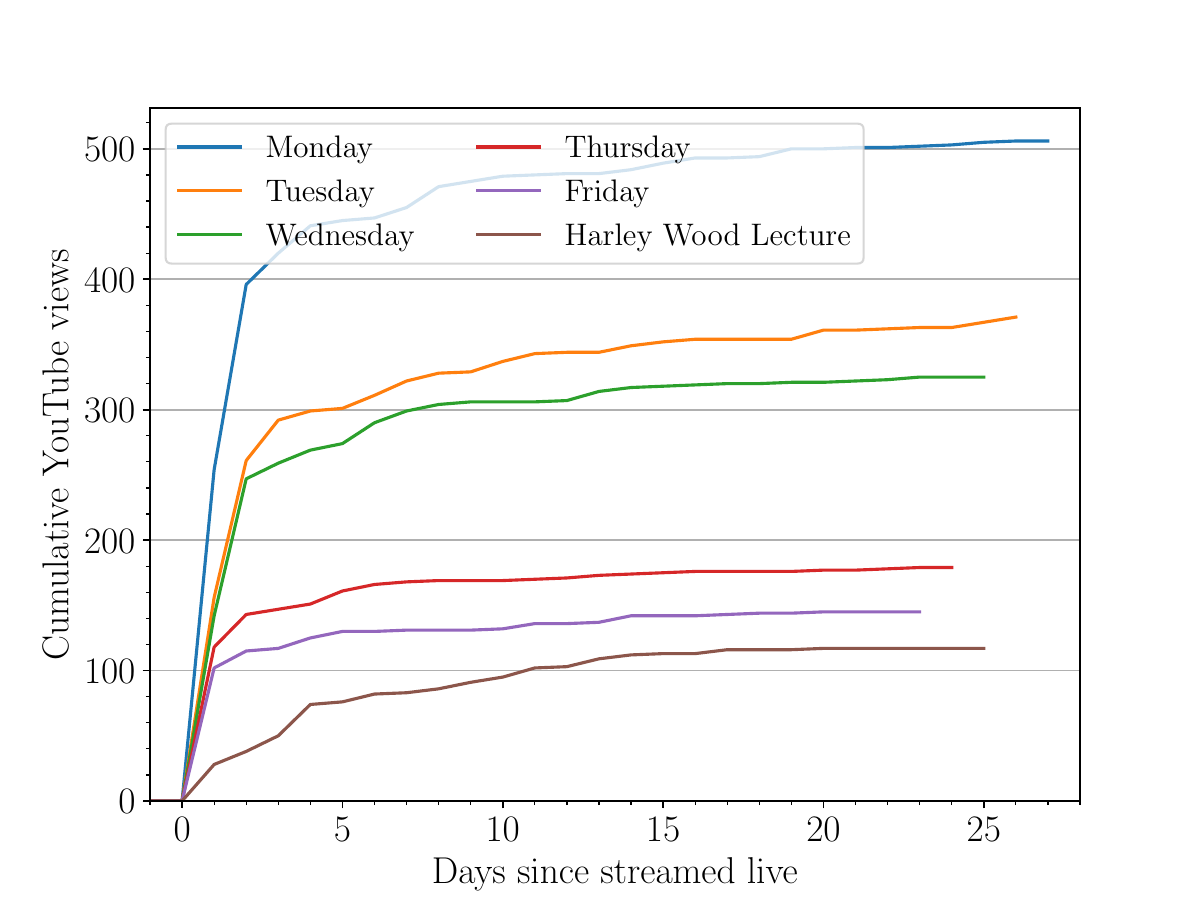}
        \caption{YouTube views.}
        \label{fig:stats:rec:view}
    \end{subfigure}%
    \begin{subfigure}{0.55\textwidth}
        \includegraphics[clip, trim = {0 0 40 40}, width = \textwidth]{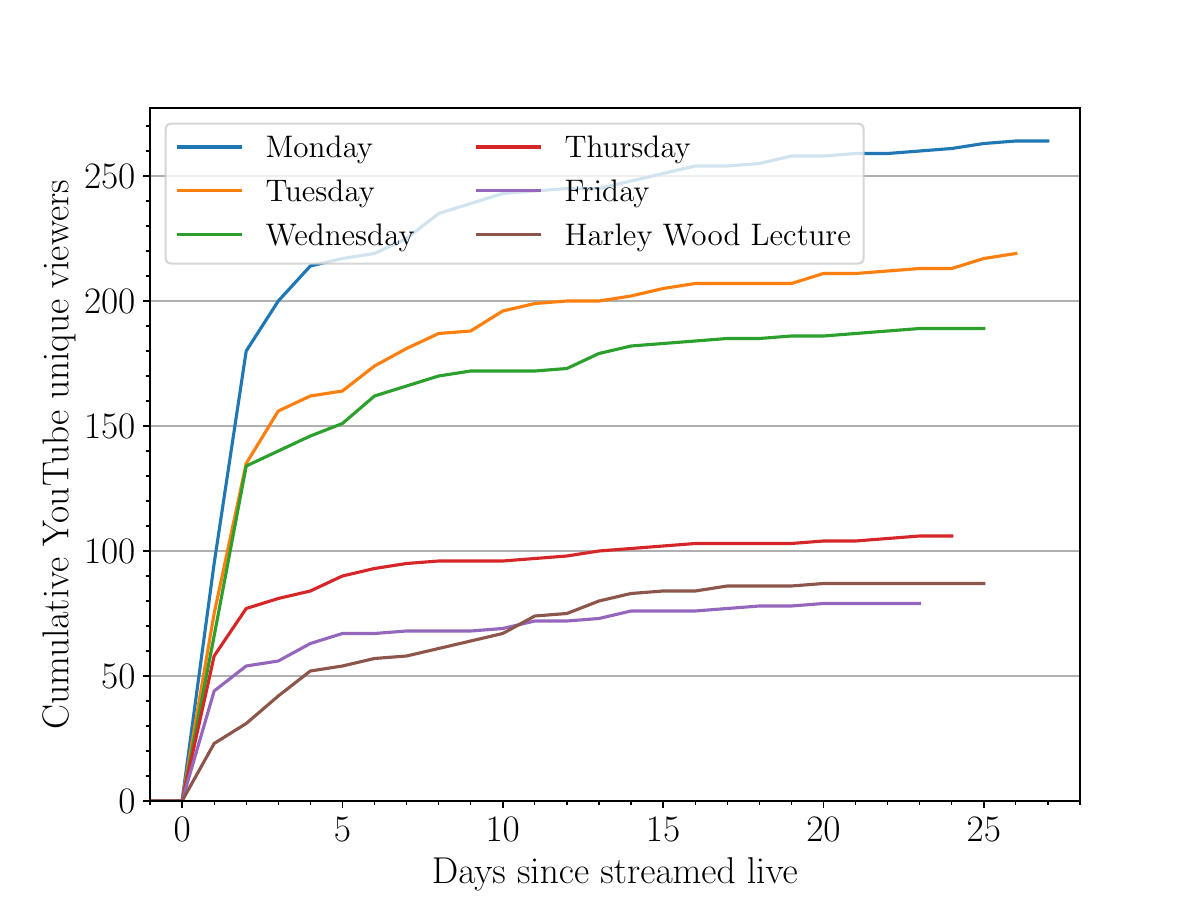}
        \caption{YouTube unique viewers.}
        \label{fig:stats:rec:uniq}
    \end{subfigure}
    \caption{
        Cumulative counts of YouTube analytics across all regular session recordings from each day.
        \emph{Left:}~Count of views.
        \emph{Right:}~Count of unique viewers.
    }
    \label{fig:stats:rec}
\end{figure*}

\section{Statistics and feedback}
\label{sec:stats}

    In this section we present and comment on the viewer counts for the Zoom meetings and YouTube streams.
    In each case we downloaded raw data from the provided analytics functionality of the respective service and reduced it using bespoke scripts written in Python.
    Separately, we sought feedback from attendees through a post-conference survey with some questions regarding aspects of the audiovisual arrangements, and we summarise the relevant feedback here.
    Throughout this section it is worth recalling that there were 34~online-only conference registrations.

    \subsection{Live attendance}

        The Zoom analytics data came in the form of a list of all participants (identified by display name, and email address if available) and the times at which they joined and left each meeting room.
        We filtered this data to remove all Adelaide staff and AV~support people, leaving only the non-Adelaide participants.
        We divided the day into 10-minute-wide bins and counted the number of unique participants who were in any meeting room for any portion of each time bin.
        As participants could be uniquely identified, this method would not double-count participants even if they were in multiple meeting rooms simultaneously, provided that their display name was the same in every room.
        The YouTube live analytics came in the form of the number of concurrent viewers sampled at each minute since the live stream began.
        We again divided the day into 10-minute-wide bins, and we set the value of each bin to the highest concurrent viewer count seen in any of the 1-minute sampling periods within each bin.
        To get the total live viewership across both streams during parallel sessions, we added together the highest concurrent viewer count from each stream.
        This method has the potential to double-count any viewers who had both streams loaded at once, but with the tools available from YouTube this effect cannot be mitigated.
        The live attendance numbers for Zoom and YouTube are presented in Figure~\ref{fig:stats:live}.

        Live attendance numbers from Zoom and YouTube are generally similar with peaks of about 10--15~attendees online during the first two days of the conference.
        Online attendance dropped lower as the week went on, possibly due to conference fatigue or the particular timetabling of more general-interest talks.
        The major differences between the Zoom and YouTube attendance figures are around the 14:00--16:00~plenary block, where Zoom attendance tended to rise higher than the morning or afternoon sessions, reaching up to 20~attendees, whilst the YouTube attendance remained consistent.
        The special sessions for the Annual General Meeting~(AGM) and Decadal Plan announcement drew additional Zoom participants, but had no discernible effect on the YouTube viewership.
        Zoom access to the AGM and Decadal Plan special sessions was made available to the full ASA membership, thus the higher attendance of this session through Zoom is expected.

        Excluding the AGM, Decadal Plan, Chapter meetings, and Harley~Wood Lecture, we counted 60~unique Zoom attendees of the regular sessions across the week.
        Of those, 29 were registered online attendees (of an expected total of 34), 17 were registered in-person attendees, and 14 were unregistered.
        As there were 194 in-person registrations, this means that nearly 10~per~cent of in-person attendees used the Zoom meetings at some point.
        These 60~Zoom attendees collectively spent 202~person-hours in regular sessions (an average of about 3.4~hours per person).

    \subsection{Recording viewers}
    \label{sec:stats:rec}

        Once a YouTube livestream ended, it automatically converted into a published video with the same behaviour as any other YouTube video.
        YouTube provides analytics for published videos, which count the viewership on each day within a queried date range.
        We analysed the number of views and number of unique viewers on each published meeting recording in the 28~days after the beginning of the conference.
        YouTube counts a ``view'' when someone watches a video for more than 30~seconds.
        If the webpage is refreshed, or the video loaded on a different device, the same user can contribute additional views, up to a maximum of 5~views in 24~hours.
        However, YouTube applies an algorithm to remove any views that are not ``legitimate'' such as deliberate attempts to artificially inflate view count.
        YouTube counts ``unique viewers'' by ``accounting for cases when a viewer may watch content on different devices or when multiple viewers share the same device''~\parencite{UniqueViewers}.
        It uses both signed-in and signed-out traffic to estimate the true number of viewers.
        Unique viewers is a useful metric to study the true audience size for a channel.

        We stacked the counts of the two regular session streams that occurred on each day to get the total viewership for each day's content (leaving the Harley~Wood Lecture separate).
        We then calculated the cumulative count of views and unique viewers for each day's content.
        The YouTube recording viewership numbers are presented in Figure~\ref{fig:stats:rec}.
        In the counts of both views and unique viewers, we see a consistent pattern of reducing attention given to the content from later in the week of the conference.
        Whilst the session recordings show a consistent pattern of accumulating the majority of their engagement within the first day, the Harley~Wood Lecture appears to break this pattern.
        The Harley~Wood Lecture has accumulated views by unique viewers more gradually.
        As the lecture is listed publicly on YouTube, this video may continue to accumulate views for a long time as it can be promoted by the YouTube algorithm.
        By contrast, the session recordings are unlisted and unlikely to garner many more views as time goes on.

        The number of unique viewers across all conference recordings (including the Harley~Wood Lecture) over the 28-day period selected is 423, controlling for the same unique viewer watching multiple different videos.
        To put this number in perspective, the ASA has 824~members as of 2025.
        Apart from the Harley~Wood Lecture, it should generally be the case that only members of the ASA have viewed the regular session recordings, so we might expect that about half of the ASA membership viewed some amount of the conference recordings.

        Another metric provided by YouTube analytics is the average view length.
        Averaged across all conference recordings, the average view length was about 10~minutes with little variation between videos.
        This suggests that viewers were typically referring to the recordings to watch specific talks, not full sessions.

    \subsection{Survey feedback}

        \begin{table*}[t]
            \centering
            \caption{Analysis of the survey questions regarding the online experience.}
            \label{tab:stats:survey}
            \begin{tabular}{r c c c}
                \toprule
                Question (rating scale 1--10) & No. responses & Mean & Std. dev. \\
                \midrule
                Standard of Zoom support      & 28 & 8.6 & 1.4 \\
                Standard of YouTube streaming & 23 & 8.9 & 1.4 \\
                \bottomrule
            \end{tabular}
        \end{table*}

        We distributed an online survey after the conference concluded.
        The survey closed with 58~responses, of which \SI{50}{\percent} were from students and \SI{43}{\percent} were academic or professional staff.
        The survey requested responses to 11 questions on a rating scale from 1~(poor) to 10~(excellent) with 5 being neutral or non-applicable.
        In hindsight, using 5 to indicate a non-applicable response complicates the interpretation of the data and another method for this should have been sought.
        The survey also included three free-response fields, for ``aspects that worked especially well'', ``did not work so well'', and other general feedback.

        Two of the questions specifically regarded the online experience.
        The prompts were ``standard of Zoom support'' and ``standard of YouTube streaming''.
        More than half of the scores for each of these questions were 5, likely from in-person attendees with no opinion on these aspects.
        Excluding all scores of 5, we calculated the mean and standard deviation of the remaining scores.
        These results are presented in Table~\ref{tab:stats:survey}.
        Both questions received mean scores above 8, which we consider to be a good result.
        Furthermore, not captured by these statistics was that about half of the responses to each question were scores of~10.

        The free-response survey questions mostly noted negative feedback on the quality of the catering\footnote{
            Further discussion of the issues with the catering is beyond the scope of this report, but will appear in a separate report tabled to the ASA.
        }.
        Of the responses that directly commented on the AV components, there was a positive observation on the utility of the recordings~(1) and that the AV in talk sessions sounded clear and ran smoothly~(4).
        There were negative observations regarding the microphone volume for both in-person and online speakers~(2; we acknowledge isolated issues with volume but do not believe there was a widespread problem), broken PowerPoint animations~(1), and a lack of online options for interacting with the poster sessions or networking~(2).
        We offer comment on some of this negative feedback in the following sections.

\section{Lessons learned}
\label{sec:lessons}

    There are a number of lessons we wish to impart to future organisers considering a hybrid conference following any of our procedures.
    We divide these lessons into two broad categories -- procedural and institutional -- depending on whether they relate to the policies and procedures we designed (and thus, could have changed) or to the interactions we had with our institution (which we had less control over and serve more as a warning in this case).
    There are also benefits to organising a conference in hybrid mode that may not be apparent from the outset, and we briefly comment on our impressions of these benefits.

    \subsection{Procedural considerations}

        \begin{itemize}
            \item Our policy of running all presentations via the venue PC did run into issues when it came to speakers with PowerPoint presentations.
            About two-thirds of the \num{\sim100}~contributions were delivered by PowerPoint, which was a far higher proportion than we expected when setting our policy.
            Issues we encountered related to fonts being missing when opened on the venue PC or animations in the PowerPoint presentations not working as the author had designed.
            There were also issues with embedded videos either not playing at all or not playing correctly.
            Some speakers came to the session with a Google~Slides link, which we did not prepare for.
            It was difficult to arrange the correct permissions to view the slides from the venue PC at short notice.
            We encouraged each author to upload a PDF backup for cases of issues with fonts, but anecdotally we found that many speakers were reluctant to rely on this backup because of the lack of animations and embedded videos.

            Due to the high prevalence of PowerPoint presentations in the astronomy community, we suggest that future conferences may wish to have a policy that can accommodate speakers delivering their presentations from their own device.
            This would also cover cases of Google~Slides or other software.
            Speakers could be given the option of joining the Zoom meeting (or equivalent) used for the session and having their own screen-share ready to go when their allocated talk time begins.
            There would be no need to connect their laptop over HDMI because the screen-share would be visible through the Zoom connection on the venue PC.
            Microphone audio would have to remain routed through the venue PC (and thus the speaker must remember to mute their laptop), and a backup option of running the presentation on the venue PC in case of technical problems would be wise.

            \item Perhaps because of the conference being hosted in winter, we had a number of members of the AV~team fall ill immediately before or during the conference.
            This included one of the two coordinators before the conference started, then the second coordinator halfway through the week, and one of the HDR volunteers.
            Between the 15~people remaining to take shifts, some had to contribute up to 10~hours of work when others were as low as 3~hours.

            Ideally, the number of HDR~volunteers should have been higher than 15 to permit a more fair division of work.
            It would also have been helpful to have a third person on the coordination team, meaning a third person with deep knowledge of all of the systems and procedures.

            \item Due to some oversight, we did not promote the Slack channels as a route for taking questions or continuing discussion after talks, despite this being our intent.
            We could have advertised this idea more clearly and Slack may have seen more take-up as a result.

            \item There was some confusion for the online attendees regarding swapping between the ``normal'' protected Zoom links that were used for most sessions, and the public Zoom links for Chapter sessions.
            Again, some clearer communication would have helped attendees to be aware that certain sessions had special links.

            \item We found that the Indico materials package download attempted to use the full title of each speaker's talk to set the name of the folder containing the materials.
            Some speakers had submitted very long talk titles, and as a result we ran into the Windows restriction that a complete file path cannot be longer than 255 characters.
            A character limit on the talk title (perhaps 100 characters) would have helped with this.
        \end{itemize}

    \subsection{Institutional considerations}
    \label{sec:lessons:institutional}

        \begin{itemize}
            \item We made the assumption that it would be easy for our institution to route the microphone audio feed in each room through the venue PC -- indeed we assumed it was already routed this way.
            We found, surprisingly, that this was not the case in one of our venues, and we were informed by IT~support that it would not be trivial to achieve with their existing equipment.
            Ultimately, we had to hire an external microphone system for that venue regardless (to get four microphones), so we sidestepped this issue from the institutional side.

            We advise starting discussions with your institution very early to find out what accommodations they can make.
            If their advice will be that an external hire is necessary, then this is best to find out with months to spare, not mere weeks.

            \item We had to request the installation of \LogiOptionsPlus on the university-managed PCs, and were surprised to find that IT~support had a projected turnaround of four~weeks while a security assessment was performed.
            After further discussion, this security assessment was fast-tracked due to the conference being only one~week away.
            However, such exceptions may not always be possible and we advise organisers to determine their software requirements as early as possible if they intend to use university infrastructure.

            \item Ultimately, using the university-managed PC in each venue had some pain points and we might have avoided using those PCs at all if we had known what to expect from the beginning.
            In addition to the strict policies regarding installation of new software (see above), these PCs were also configured not to store any user preferences.
            This led to a lot of repetitive work for the AV~team as it was necessary to reconfigure \LogiOptionsPlus and Zoom at the start of every session.

            Of course, in one of our venues this PC was the only route to making use of the microphone audio, but in the other venue we could have borrowed or hired a different device to take the microphone audio and connected it to the projector over HDMI.
            Despite the downsides, there were advantages to using the venue PCs, such as a gigabit ethernet connection and the convenience of the monitors, mouse, and keyboard already present.
        \end{itemize}

    \subsection{Flow-on benefits of a hybrid mode conference}
    \label{sec:benefits}

        \begin{itemize}
            \item The session recordings on YouTube were a by-product of the decision to live stream the sessions.
            As noted in Section~\ref{sec:stats:rec}, these recordings were well-utilised by attendees as well as the wider ASA.

            \item Our policy of running all presentations through the venue PC, which was done mainly for the benefit of a consistent Zoom screen share, forced a streamlined process for swapping between speakers' presentations.
            This certainly made the changeover between talks much faster than allowing each speaker to plug in their own laptop over HDMI, configure their display settings, and put the presentation into fullscreen on their own.

            \item Our policy of running all presentations through the venue PC also forced a situation where every speaker uploaded their slides to the conference platform, which otherwise may not have happened.
            Having the slides available, similar to having a session recording, is a useful way to disseminate the science further to attendees who could not make it to a particular session.
        \end{itemize}

\section{Conclusion}
\label{sec:conclusion}

    In this report we have presented an overview of all aspects of the audiovisual setup for a national-level academic conference arranged with hybrid attendance options.
    Our objective when arranging the conference was to make as much use of our existing institutional resources as possible, so the steps we have outlined here are not intended to be prescriptive of a single or ``best'' way to arrange such an event.
    Still, we feel our experience is worth considering as a demonstration of what can be possible, with the appropriate institutional support.
    We would encourage other conference organisers to use this report as a starting point and make variations according to their own technical capabilities and resources.
    A hybrid conference need not be an expensive technical undertaking, and there are benefits which go beyond the online attendees, to both the in-person attendees and the wider scientific community.
    We believe these are worth enabling wherever possible.

\section*{Acknowledgements}

    The organisers thank Vanessa Moss for constructive discussions about the Zoom meeting and YouTube livestream configurations when preparing for the conference, and feedback on the content of this report.

\sloppy
\printbibliography[heading = bibintoc, title = References]

\end{document}